# Wafer bonding solution to epitaxial graphene – silicon integration

*Rui Dong[1], Zelei Guo[1], James Palmer[1], Yike Hu[1], Ming Ruan[1], John Hankinson[1], Jan Kunc[1], Swapan K Bhattacharya[1], Claire Berger[1,2] and Walt A. de Heer[1*]*

*[*] Corresponding Author,*
[1] School of Physics, Georgia Institute of Technology, Atlanta GA, 30332, USA
[2] CNRS-Institut Néel, BP166, Grenoble Cedex 9, 38042, France

The development of graphene electronics[1,2] requires the integration of graphene devices with Si-CMOS technology. Most strategies involve the transfer of graphene sheets onto silicon, with the inherent difficulties of clean transfer[3-5] and subsequent graphene nano-patterning that degrades considerably the electronic mobility of nanopatterned graphene[6,7]. Epitaxial graphene (EG) by contrast is grown on an essentially perfect crystalline (semi-insulating) surface, and graphene nanostructures with exceptional properties[8-11] have been realized by a selective growth process on tailored SiC surface that requires no graphene patterning[9,12,13]. However, the temperatures required in this structured growth process are too high for silicon technology. Here we demonstrate a new graphene to Si integration strategy, with a bonded and interconnected compact double-wafer structure. Using silicon-on-insulator technology (SOI)[14-16] a thin monocrystalline silicon layer ready for CMOS processing is applied on top of epitaxial graphene on SiC. The parallel Si and graphene platforms are interconnected by metal vias. This method inspired by the industrial development of 3d hyper-integration stacking thin-film electronic devices[17,18] preserves the advantages of epitaxial graphene and enables the full spectrum of CMOS processing.

Figure 1 is an illustration of the monolithic integration of both Si and SiC devices onto the same double wafer, showing CMOS devices patterned on the thin crystalline Si wafer on top, graphene transistors on the SiC wafer below, and metallic vias patterned through the Si wafer for 3d interconnection between the two electronic platforms. This contrasts with most Si/graphene integration schemes[19-21] where graphene- and Si-device areas are implicitly designed side by side on the same plane. The Si wafer transfer solution described below in detail presents several advantages. The transfer can be realized in principle on the wafer scale (Si to SiC transfer at the wafer has been already realized[22]) and the resulting double-wafer is compatible with silicon-VLSI. The top monocrystalline Si surface present the quality required for CMOS, that was difficult to obtain by growing Si on SiC by chemical vapor deposition, molecular beam epitaxy or electron beam evaporation[22]. The transfer relies on Si to EG/SiC wafer bonding that is based on the silicon-on-insulator (SOI) technique, a mature industrial process in silicon technology. In our case for Si to EG/SiC bonding we have adapted the process by adding an $Al_2O_3$ layer to assist bonding. Epitaxial graphene is grown on the crystalline SiC wafer[13] prior to the Si-SOI transfer, therefore the high temperature graphene on SiC growth is not limited by the lower Si melting point, allowing very good quality (nanostructured) graphene, and any post-processing if required. Moreover, the graphene layers/nanoribbons remain untouched on their growth substrate. This ensures that graphene's integrity, interface and nanostructure properties are preserved. Moreover, access to the graphene structures from above provides significant architectural flexibility for graphene device interconnects. Finally, the often-quoted[23] drawback of the epitaxial graphene is the SiC substrate cost (currently about $20/cm$^2$ and decreasing) that deserves to be addressed upfront. Considering, that high-end consumer electronics processors currently cost more than $1000, it is clear that if a SiC substrate were to be used in those, the SiC cost would amount to *only a few percent* of the total price, which is very reasonable, especially if unsurpassed performance is achieved.

Figure 2 shows a process flow of the proposed Si to EG/SiC integration. (1) silicon oxide is grown by thermo-oxidization on a commercial monocrystalline Si wafer. (2) Hydrogen ions are implanted in the oxidized-Si wafer. (3) 30 nm thick aluminum oxide is deposited by atomic layer deposition on the $SiO_2$/Si dies (4-5) EG is grown on SiC. Non graphene covered areas are managed on the wafer, either growing submonolayer EG on the C-face, or by plasma etching graphene in patterned area, or by growing graphene only on the sidewalls of trenches etched in 4H-SiC (Si-face). (6) 30 nm ALD-$Al_2O_3$ is deposited on EG/SiC. Because of growth selectivity, $Al_2O_3$ growth is confined in SiC regions not covered with graphene. (7) The $Al_2O_3$/$SiO_2$/Si and $Al_2O_3$/EG/SiC wafers are bonded together using $Al_2O_3$ as a bonding interface. (8) Upon heating the bonded wafers to (400˚C), the Si wafer splits at the ion implantation depth (smart-cut), leaving a thin monocrystalline Si layer bonded to the EG/SiC wafer. (9) Windows are opened by standard microelectronic patterning and etching processes to expose some area of the buried EG layer. (10) EG and the top crystalline silicon layer are interconnected by metal pads. This process can clearly be generalized to wafer size (SiC wafers are now commercially available up to 150mm diameter). We next discuss some of the process steps in more detail.

One of the key steps is the Si to EG/SiC wafer bonding (step 7). Si-wafer size bonding has been an industrial process for two decades[24], but there are only few reports on SiC wafer to Si wafer bonding [22, 25-27], and none of Si on graphitized SiC. The primary challenge was to realize bonding to the SiC substrate coated with graphene that is well known for its non-sticking properties. Our solution consists of adding an intermediate alumina layer between the Si wafer utilizing graphene free regions of the SiC wafers. This solves also two of the mains challenges of wafer bonding. One is the stress during thermal treatment because of the different thermal expansion coefficients between Si and SiC. The second is that the two facing surfaces have to be smooth and flat. Significant SiC surface step bunching during EG growth can be a limiting factor.

Figure 3a shows an optical view of several bonded 3.5mm x 4.5mm samples (Si/SiO$_2$/Al$_2$O$_3$ - Al$_2$O$_3$/EG/SiC). Gold color indicates strong bonding contrasting with weaker bonding in the blue (or green) areas that are located mostly at the sample edge. Figure 3b shows the optical image of the 2 halves of a bonded wafer after smart-cut splitting (step 8 above). On the left is the SiC die with the Si layer bonded to it (Si/SiO$_2$/Al$_2$O$_3$-Al$_2$O$_3$/EG/SiC stack). The darker area is where crystalline Si has transferred from the Al$_2$O$_3$/SiO$_2$/Si wafer shown on the right. The shape of the transferred silicon layer (left) matches precisely the bright area on the Si wafer die (right), which shows the success of the smart-cut transfer. The profilometer scans of Figure 3d on the transferred wafer (black trace) and on the Si wafer (red trace) wafers show that in this example a Si/SiO$_2$ layer 1.2 μm thick was transferred.

The successful Si smart-cut transfer shown in Fig 3a-b demonstrates the wafer bonding strength. The wafer splitting is caused by the formation of molecular hydrogen blisters at the specific depth of proton implantation in the Si wafer. The SiC/Si wafer bond needs to be sufficiently robust to withstand the stress of the smart-cut process. It should be noted that bonding of small wafer dies like those used here (3.5x4.5mm$^2$) is particularly challenging and requires much higher bonding energy and much cleaner interfaces than for wafer scale bonding. For instance, for a 4-inch Si wafer, particles as small as 1 μm diameter typically result in a 5 mm diameter unbonded area[28], which is the size of SiC dies. Therefore thorough cleaning is required: contaminant particles, mostly found at the edges due to dicing and handling must be removed. Figure 3c is a scanning electron microscope (SEM) image of the bonded interface between transferred silicon and SiC. The image is taken with a tilt angle at the edge of the Si layer and shows the section of the SiO$_2$ coated Si bonded to Al$_2$O$_3$/SiC. The image shows that the interface is clean and sharp with no gaps or cracks.

The Si wafer transfer method proposed here preserves the structural quality of EG. A key point in the process is to selectively grow alumina at specific locations by atomic layer deposition (ALD) (step 6). In the process alumina selectively coats the prepared graphene-free regions (that are obtained by growing sub-monolayer graphene or by removing locally graphene by plasma patterning). The selective coating is realized by depositing ALD –$Al_2O_3$ directly with no pre-seeding, in contrast to the deposition of dielectric for graphene field effect transistors where special treatments are use to force $Al_2O_3$ to cover graphene (see for instance [29, 30]).

In the example of Figure 4, sub-monolayer graphene was grown on the C-face of 4H-SiC. Raman spectroscopy is used to identify graphene regions (characteristic 2D and G peaks, see for instance Fig. 4c) from bare SiC. Fig. 4a shows an AFM image of the surface after ALD-$Al_2O_3$ direct deposition. The dark area is a single layer EG layer draped over the SiC substrate steps. The graphene layer is recognized also by its surface pleats (white lines) as is usual for graphene on the C-face [9, 13]. As is clear from the AFM image graphene is clean from alumina. Alumina coats preferentially the surrounding bare SiC substrate, as shown by the surface roughness contrasting with that of graphene (AFM line profile of Fig. 4b). Here we use to our advantage the non-wetting properties of graphene, that is in general problematic when growing dielectric on graphene for top-gating (a functionalized or seed layer is required [29, 30]). As alumina is deposited, the uncoated graphene becomes lower than the $Al_2O_3$-coated SiC. This prevents EG from making direct contact with the Si wafer die in the following bonding step because bonding happens only between the $Al_2O_3$ coated areas. The Raman spectra of EG/SiC (Fig. 4c) show that the characteristic G and 2D peaks of graphene remain unchanged before and after ALD-$Al_2O_3$ deposition and no D peak indicating of disorder is seen in either case.

The successful bonding indicates that graphene is not involved in the bonding process (graphene on the contrary delaminates easily). In order to connect the top (Si) and bottom (graphene) electronic layers, openings are etched in the bonded Si wafer, dry and wet etching is used to open the vias for metallic 3d connection between the Si and graphene layers. The Raman spectrum of Fig. 4d shows that graphene is not significantly affected by the optimized etching process used to open the large windows of Fig. 4a-b through the $Si/SiO_2/Al_2O_3$ layer (etching will certainly be further optimized as the process develops). This result is confirmd by transport data below (Fig. 4c) As seen in Fig.S1d: very low or no Raman D peak was observed after etching for multi-layer graphene on two different locations indicated by the green dots on the optical image. Note that the etching time is adapted to the thickness of the crystalline Si transferred. The $SiO_2$ "mask" was removed by a short buffered oxide etching (BOE) at room temperature (see methods section below for details).

In this study, successful Si wafer die bonding has been realized on two types of EG samples: C-face SiC substrates coated with a sub-monolayer graphene layer and on an array of nanoscopic graphene ribbons grown by the templated growth method [9, 12] on the Si- face, as demonstrated now. Figure 5 shows Si to structured EG/SiC integration. As can be seen in the optical image of Fig. 5a-b, successful bonding is obtained between Si wafer die and structured EG/SiC. In this example arrays of 200 parallel graphene ribbons (100nm x 100 μm) were selectively grown on the sidewalls of trenches patterned in the 4H-SiC substrate (Si face)[9, 10, 12, 13]. The 50nm deep vertical trenches dry-etched in SiC (Fig. 5c) recrystallize into well-defined crystallographic facets upon annealing around 1500°C resulting in 100nm wide sidewall templates. Because graphene growth rate is slower on the Si (0001) face, graphene ribbons are first formed on the sidewall facets. By adjusting the growth conditions and time, ribbons can be selectively grown, as seen in the electrostatic force microscopy (EFM) image of Fig. 5d. It is important to note that in this case graphene nano-structuring is realized *prior* to substrate bonding. There is therefore no temperature limitation to produce high quality, smooth edged graphene nanostructures. It was also demonstrated that sidewall graphitization is not limited to lines and the etched SiC substrate acts as a template for graphene growth[9, 13].

The main goal of the Si to graphene integration is to interconnect the graphene device platform to the Si-CMOS technology on the same wafer (steps 9 and 10). Figure 6a-b show an example of the proposed integration. Windows (20µm side) were etched in the top $Si/SiO_2/Al_2O_3$ layer by a combination of standard dry and wet etching to partially expose a 4 $\mu$m wide and about 30 $\mu$m long EG area grown on the C-face. The EG area, shown by the white dashed contour in Figure 6a, lies partly underneath a 1 µm thick monocrystalline silicon layer. Eight evaporated metal strips (Ti/Pd/Au : 0.5 nm/20 nm /50 nm) are prepared by conventional lithography and lift-off techniques and connect the bottom EG to the top Si wafer die where the pads extend for electrical measurements.

The resistance measurements below confirm the Raman data after ALD deposition and window etching that the characteristics of graphene are not affected by the process. From the resistance measurements several conclusions can be drawn. (i) The metal leads are continuous from EG to the top Si surface, as is also observed from the tilted view on Figure 6a. (ii) Graphene is not disrupted by the bonding process. A finite resistance of a few hundreds ohms is measured between any 2 leads, as shown in Figure 6c. (iii) Exposed and Si covered graphene have a similar resistivity $R_{sq}$ = 200-300 $\Omega$/sq, typical for highly doped single or few layer graphene[31, 32], and a maximum contact resistance $R_C$~600 $\Omega$.µm, which is in the range of published values for metal to graphene contacts[33]. The graphene quality and good metal connection to the top silicon wafer die have been further tested by applying a large current through the leads. The IV characteristics are linear and current density, as high as 1.5mA/$\mu$m, can be reversibly applied on leads connecting Si-covered and exposed graphene, with no observable degradation of the leads or of graphene.

We have demonstrated here the critical step of a graphene – silicon integration scheme to produce a monolithic integration of two wafers acting as interconnected parallel electronic platforms. The process is quite flexible and we envision the development of electronic devices on both platforms. CMOS technology can be implemented on top of the silicon wafer, which surface is entirely free for device processing. The smart-cut technique[24] allows to choose the thicknesses of the Si layer (5 nm to 1.5 µm) and of the $SiO_2$ oxide (5 nm to typically 5 µm). Ion implantation, epilayer growth and standard lithography techniques can be safely implemented to the top Si layer, and even more so when the graphene is protected during processing, i.e. if the windows or vias are fabricated as the last step. Epitaxial graphene is in any case very robust to chemical treatments (Figure 4d). Moreover EG on SiC can safely withstand temperatures up to 450°C in air and 1000°C in vacuum, since these annealing steps are used routinely to clean graphene from contaminants (as demonstrated in the AFM image and Raman spectra of Figure 7). The effect of air annealing on graphene is shown.

These studies show that fully developed graphene devices and interconnects on the SiC surface can be produced prior to bonding and that they survive the bonding process. Particle contamination was the main impediment to successful monocrystalline substrate bonding in our case. However, this study was done with small dies (~15 $mm^2$), in a non-stringent clean-room environment. Despite these drawbacks, the successful bonding achieved here together with the large scale device integration demonstrated for epitaxial graphene [12, 30] indicates that this process has an industrial potential. Compared to graphene transfer or printing, this graphene to Si integration method takes full advantage of the crystallinity of the substrate and of epitaxial growth process (continuous high quality 2D sheet, well defined and reproducible interface, well known industrial grade substrate, no potentially damaging transfer required). Beyond graphene for electrodes, this integration is envisioned for high performance electronics for instance in ultra high frequency electronics [29, 30], spintronics[34], optoelectronics. We have indicated that graphene sidewall nanoribbon arrays can be integrated to Si with the same process. We believe that the recently discovered exceptional electronic and transport properties [8, 10, 11] of sidewall graphene ribbons grown directly on SiC[8, 9, 12] will become an important direction for nanoscale electronics.

In conclusion, we have developed a unique monocrystalline silicon transfer method to fabricate monolithic integration of graphene on SiC /silicon 3d stacked layers, that is fully compatible with VLSI technology and preserves graphene integrity and nano-structuring. Instead of the conventional graphene transfer technique, thin monocrystalline silicon layers are transferred onto EG/SiC wafer dies using well-established SOI wafer bonding and smart-cut techniques. The transferred crystalline silicon layer can serve as the basis of silicon-CMOS devices, and is connected to EG layer by metallic leads. High quality graphene nanostructures grown at high temperature are integrated with no degradation.

**Methods**

(1) A 300nm thick oxide was grown by thermo-oxidization on a p-doped ($10^{15}$ cm$^{-3}$) Si wafer.

(2) Hydrogen ions (140 keV, dose $8.5 \times 10^{16}$/cm$^2$) were implanted in the Si wafer at depth of 900nm, according to the implantation simulation (TRIM package). The temperature (15˚C) was controlled during implantation to avoid wafer blistering.

(3, 6) For bonding, 30nm $Al_2O_3$ was deposited directly by atomic layer deposition in a Savannah 100 ALD system, at 160˚C, using TMA as a precursor. No graphene seeding layer was used, contrary to graphene transistors, such as in refs. [29, 30].

(4-5) Submonolyer graphene was grown on the C-face of insulating 4H SiC by the confinement controlled sublimation method[13] at 1500°C. For the ribbon array, patterned SiC (Si-face) trenches were etched in $SF_6/O_2$ plasma, using Poly (methyl methacrylate) (PMMA) as a mask. After CCS growth at 1450°C, the 50nm deep sidewalls recrystallize at 29 degree from the (0001) orientation, providing a 100nm wide facet for ribbon growth. Raman spectroscopy and EFM clearly identifies graphene on the sidewalls.

(7) After $Al_2O_3$ deposition, samples were stored in DI water for more than 24 hours in order to improve their hydrophilic properties. The wafers dies were first bonded in DI water to avoid particle contaminants from air, then transferred to a pressure module. Stronger bonding strength is achieved by subsequent annealing.

(8) The bonded dies were heated up to 400°C in air so that the resulting $H_2$ pressure splits the Si wafer dies along the H implantation plane. For this, a fast ramping (10°C/min) from room temperature to 300°C was followed by a slow ramping (5°C/min) from 300C to 400°C. The bonded dies were kept at 400°C for 60 min, then naturally cooled down to room temperature.

(9) Windows in the $Si/SiO_2/Al_2O_3$ stack were opened with dry and wet etching after patterning a 1μm thick photoresist layer (Microposit SC1813) used as the dry etch mask: $SiO_2$ and Si were respectively dry etched in a $CHF_3/Ar$ RIE, and in $SF_6/O_2$ plasma. $Al_2O_3$ was removed in a solution of $H_3PO_4$: $H_2O$ (1:3) at 60°C. For the sample of Fig. 4d, the following etching recipe was used. Si was etched in $SF_6/O_2$ plasma and $SiO_2$ was etched in a $CHF_3/Ar$ RIE chamber. A shorter plasma etching recipe was used so that about 100 nm $SiO_2$ can be preserved and used as a "mask" to avoid plasma damage to the graphene underneath. The sample was further etched in a solution of $H_3PO_4$: $H_2O$ (1:3) to remove the $Al_2O_3$ residues at 60°C.

**Acknowledgements**

This material is based on research sponsored by DARPA/Defense Microelectronics Activity (DMEA) under agreement number H94003-10-2-1003, National Science Foundation under Grant No. DMR-0820382 , AFSOR and the W.M. Keck foundation. The United States Government is authorized to reproduce and distribute reprints for Government purposes, notwithstanding any copyright notation thereon.

**Figure Captions**

**Figure 1**. Illustration of a silicon-on-EG/SiC monolithic wafer integration, showing CMOS technology on a Si thin wafer on top (grey layer) and graphene devices below (blue layer). The two electronic platforms are interconnected vertically by metal vias. There is no limitation a priori on the integration design on either platform.

**Figure 2**. Process flow of silicon and EG/SiC integration: (1-3) $H_2$ implantation and $Al_2O_3$ deposition on Si; (4-6) Epitaxial growth and patterning, and $Al_2O_3$ deposition on SiC; (7) wafer die bonding;(8-10) smart-cut and metal vias fabrication to connect the top CMOS ready Si layer to the buried graphene.

**Figure 3.** Demonstration of Si on EG/SiC wafer die bonding. In this case graphene was partially grown on the C-face of SiC. (a-b): Optical images of three 3.5mm x 4.5mm wafer die Si-on-EG/SiC; golden/purple color corresponds to the bonding areas. (a) after bonding; (b) after smart cut; (left) Si on EG/SiC substrate and (right) Si wafer die showing the trace of the removed Si layer. (c) SEM images of Si-on-EG/SiC sample. The image shows a cross sectional view of the sharp and clean interface between transferred Si-$SiO_2$ and the flat SiC substrate that is partially covered by $Al_2O_3$. (d) Depth profile on both wafer dies in (b) showing that the transferred Si layer is 1.2µm thick.

**Figure 4.** (a) AFM images of a partially graphitized epitaxial graphene on the C-face, after ALD $Al_2O_3$ deposition. (Scale bar, 5µm). The dark area is bare graphene that drapes over the SiC steps (b): AFM height profile along the dotted line in (a), showing a increased roughness on the $Al_2O_3$ coating compared to graphene. (c) Raman spectra of the graphene area in (a) before and after $Al_2O_3$ coating, showing high graphene quality (no D peak). The SiC Raman peak contribution is subtracted. (d) Raman spectra of the two graphene area in the window opening after bonding and etching. The spectrum were taken at the green dots in the optical

image in the inset; Very small or no D peak is observed. The noisy spectrum between 1500 and 200 cm$^{-1}$ is due to the imperfect subtraction of the SiC Raman contribution.

**Figure 5.** Si to structured EG/SiC wafer die bonding. Arrays of 200 parallel graphene ribbons (100nm x 100 μm) are grown on the sidewalls of trenches patterned in SiC-Si face before wafer die bonding. (a) Optical image of a 3.5 mm x 4.5 mm Si-on-structured EG/SiC wafer die; The purple color indicates bonding (b) Optical image of the graphitized array seen through the SiC substrate after bonding, indicating that the bonding doesn't damage the patterned structure. (c) AFM topographic image of the array of trenches patterned in SiC, after graphitization and prior to wafer die bonding, and AFM height trace (white trace– full amplitude is 50 nm). (d) Electrostatic Force Microscopy image of a similarly prepared sample showing the contrast between SiC (dark) and the 40nm wide graphene nanoribbons (light).

**Figure 6.** Scanning Electron Microscopy images of Si on-EG/SiC substrate. Openings are provided in the Si top layer to expose buried graphene and metal pads that connect the top Si wafer die to graphene. (a) top view. The graphene area is outlined with the dotted line (green), the pas are outlined in yellow, and the monocristalline Si in red. (b) tilted view with multiple windows opened in Si to expose epitaxial graphene. Scale bars: 5 $\mu$ m (c) room temperature resistance between any two pads in (a), showing that the same resistivity (proportional to the local slope R vs distance) is measured for exposed and buried (under the central pads) graphene

**Figure 7.** Effect of annealing at 400°C for 30 minutes in air. Raman spectroscopy of a multilayer epitaxial graphene (MEG) sample before and after annealing in air at 400°C, showing the characteristic 2D and G Raman peaks of graphene. The SiC substrate Raman spectrum was subtracted. Note that the graphene 2D peak has a single Lorentzian shape as typical for MEG[9]. The extremely small D peak reveals the high structural quality of MEG

that is not affected by the annealing in air. The AFM images in the inset (scale bar: 1µm) show a patterned MEG graphene cross, before (left) and after (right) 400°C annealing in air. The white dots are residues from the resist used for patterning. The graphene cross is cleaner after anneal, that doesn't visibly change graphene. The same white line (graphene pleats) are observed and the roughness on graphene decreases from 1nm (before) to 0.1nm (after) annealing. Note that the SiC outside the graphene cross remains quite contaminated.

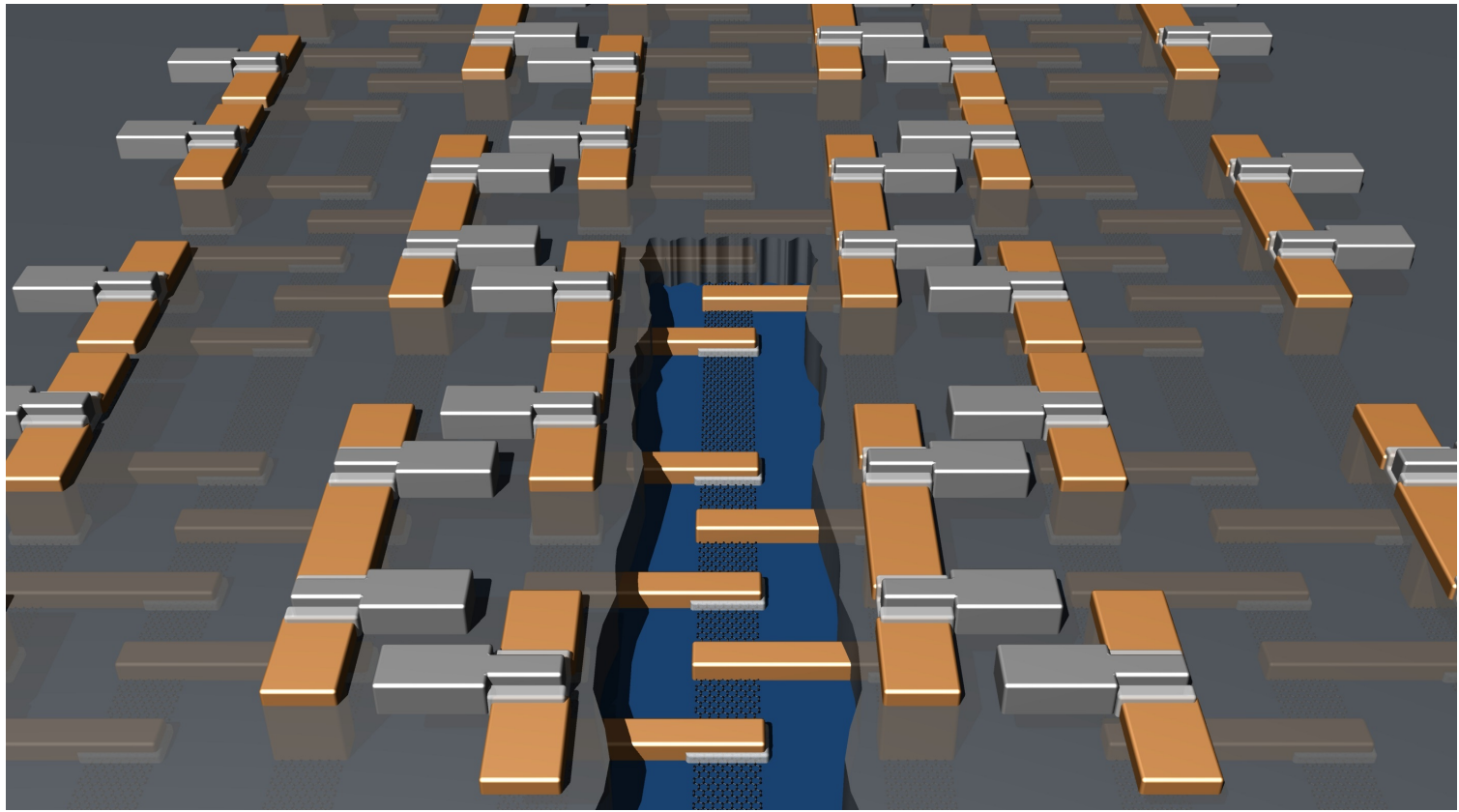

Figure 1

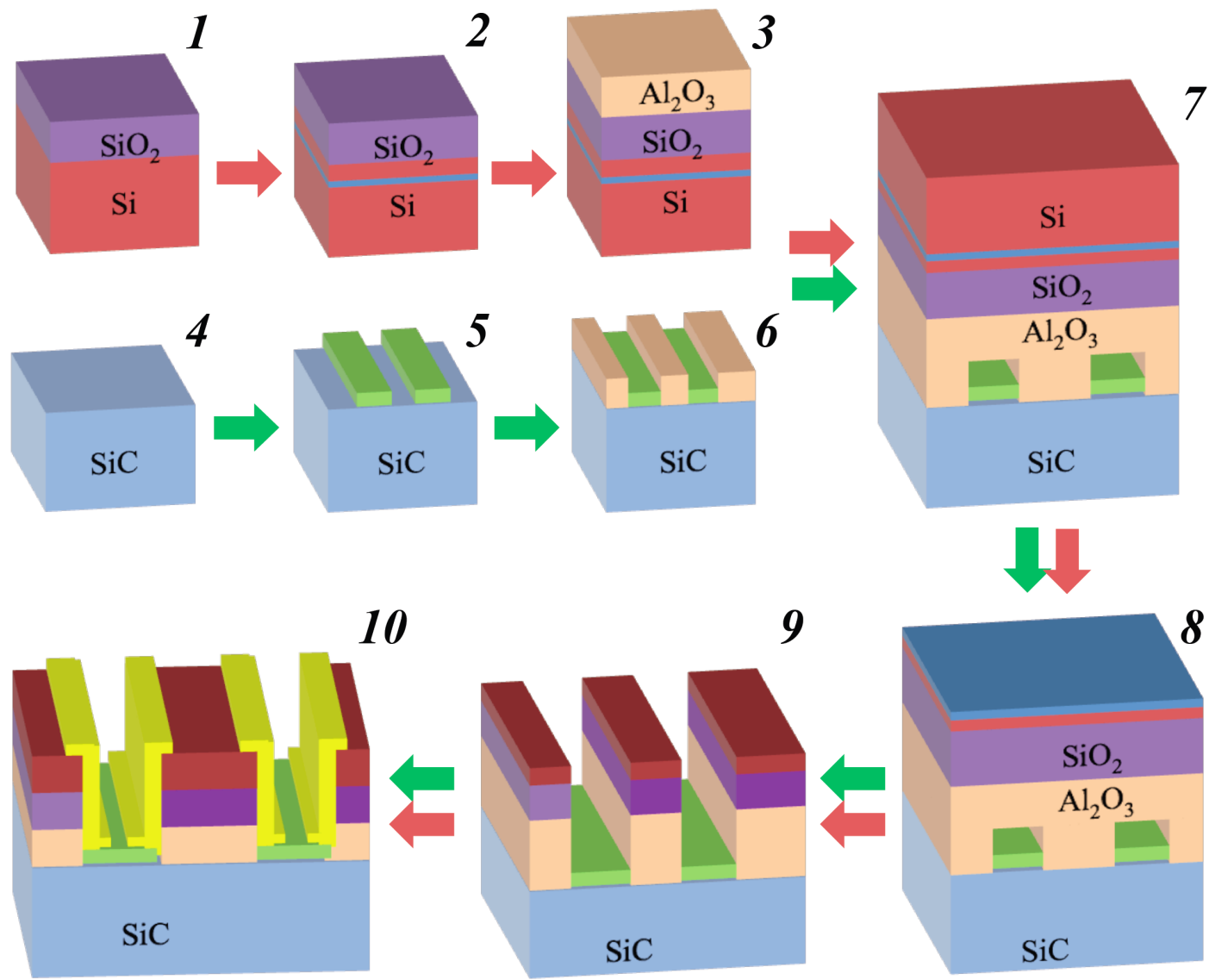

Figure 2

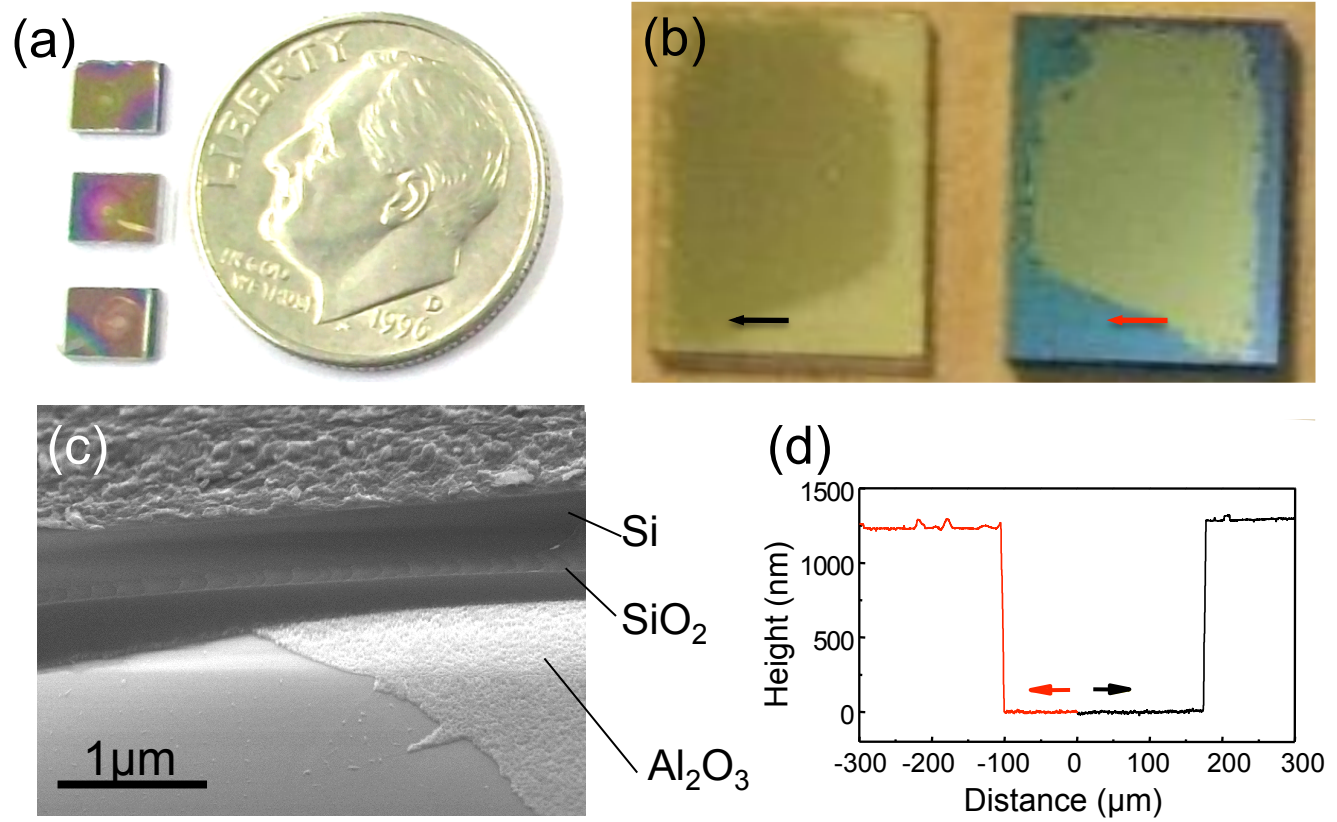

Figure 3

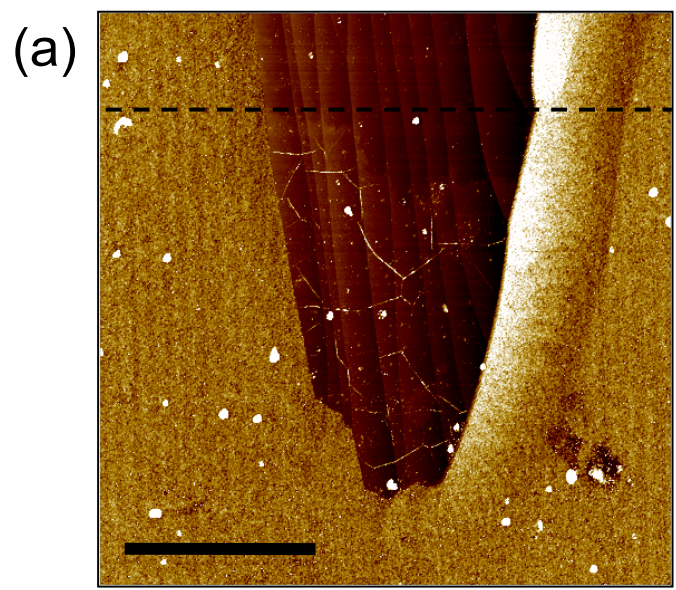
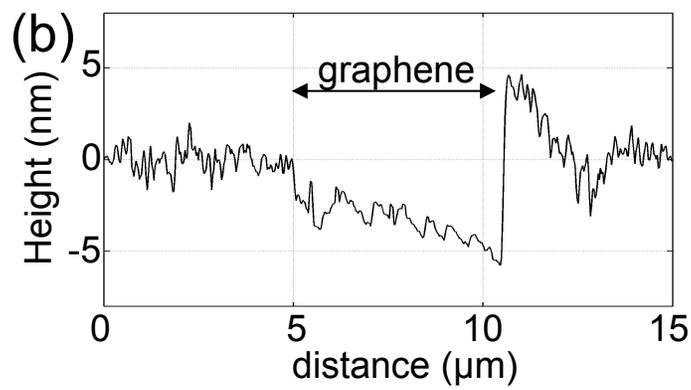
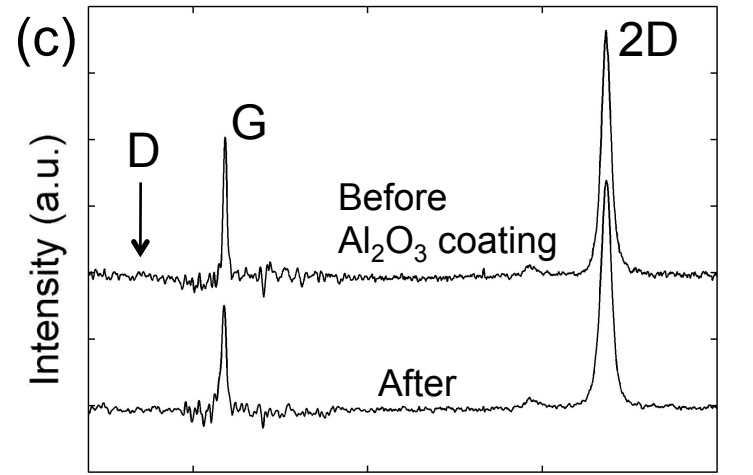
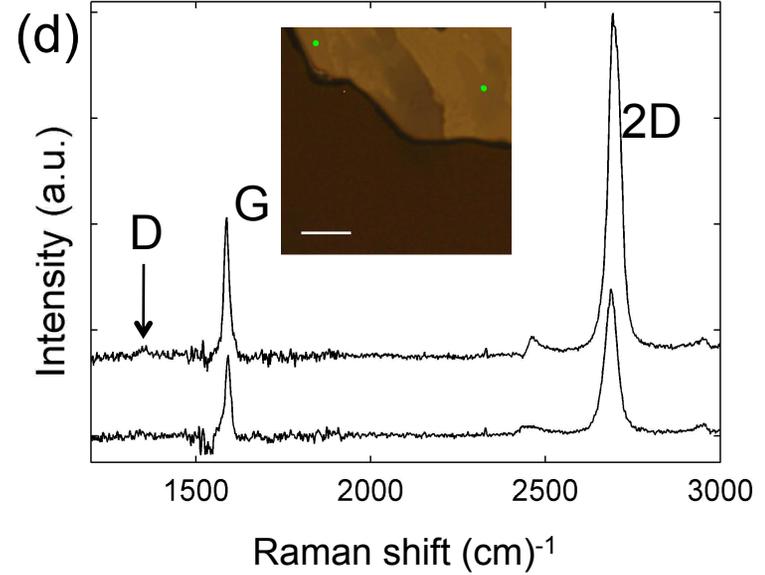

Figure 4

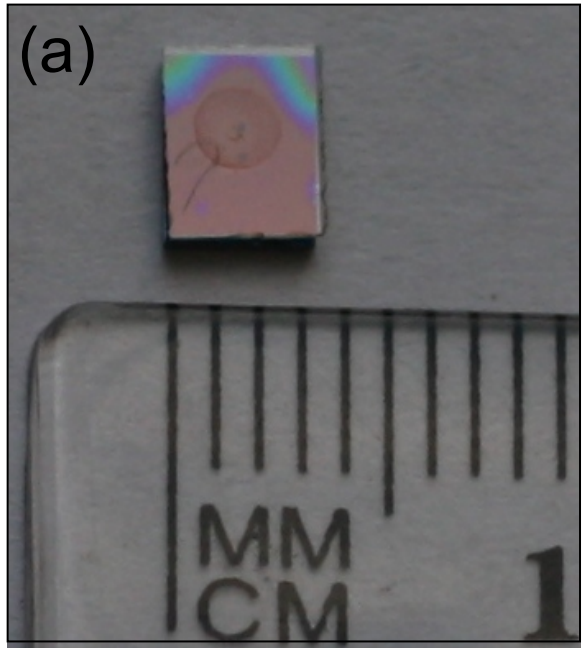
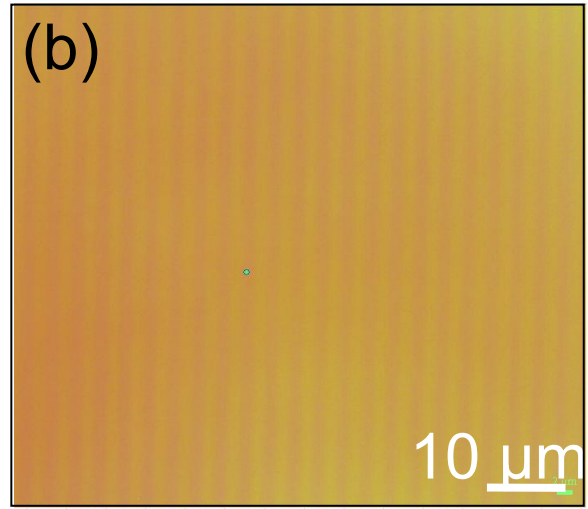
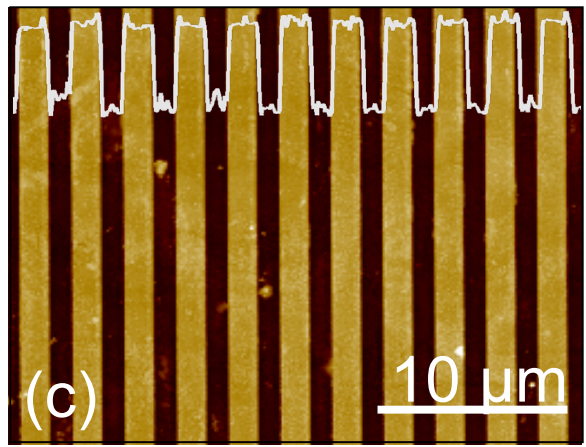
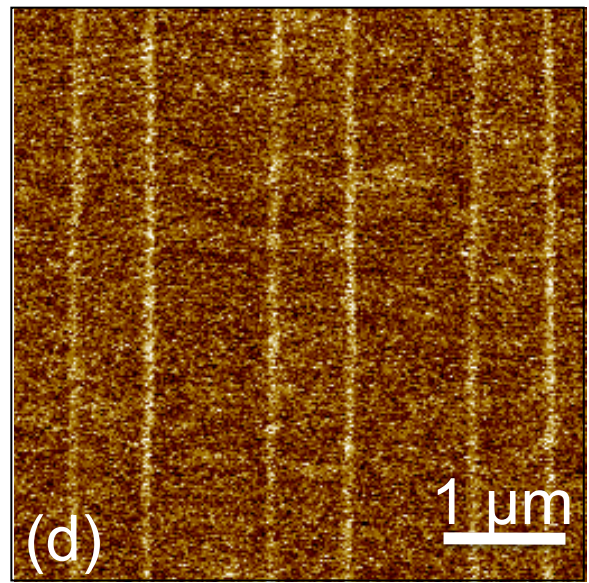

Figure 5

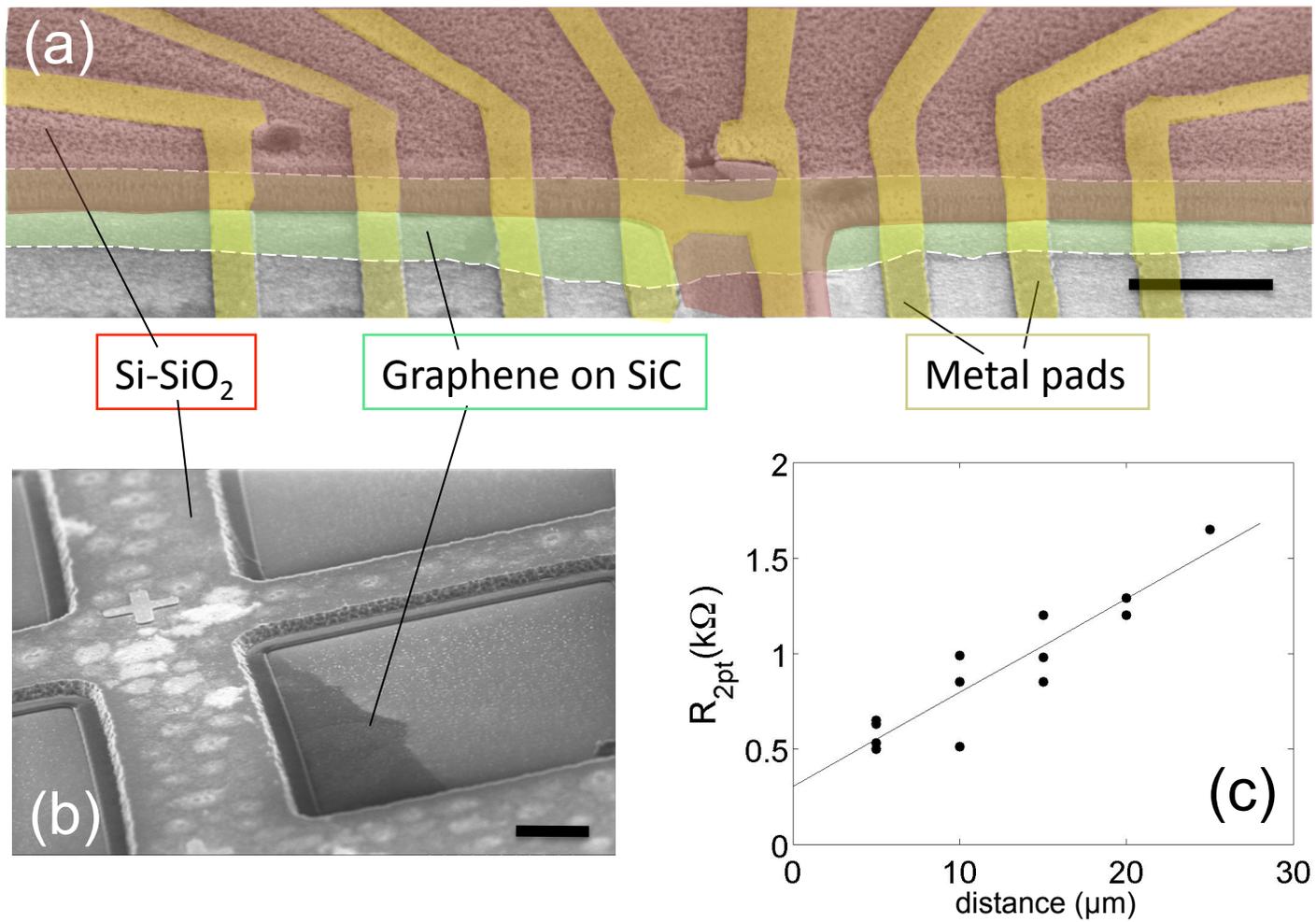

Figure 6

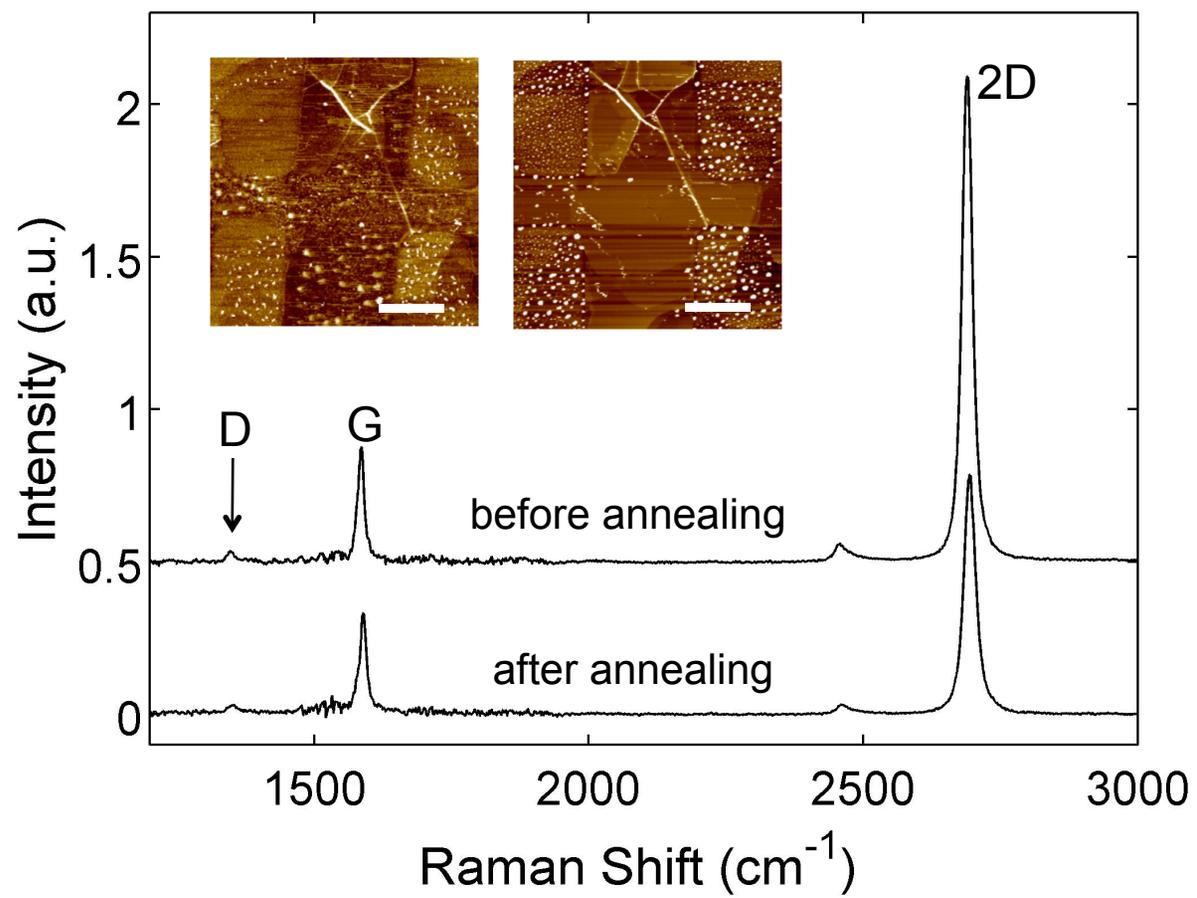

Figure 7